\documentclass[A4,twoside,12pt,reqno]{article}

\usepackage[super,compress]{cite}
\usepackage{graphicx}
\usepackage{float}
\usepackage[latin1]{inputenc}

\usepackage{amssymb}
\usepackage{amsthm}
\usepackage[tbtags]{amsmath}
\usepackage{doc}
\usepackage{latexsym}
\usepackage{amscd}
\usepackage[frame,cmtip,arrow,matrix,line,graph,curve]{xy}
\usepackage{graphics}
\usepackage{epsfig}
\usepackage{eucal}
\usepackage{mathrsfs}
\usepackage[margin=1.3in]{geometry}

\begin{document}
\thispagestyle{plain}
 \markboth{}{}
\small{\addtocounter{page}{0} \pagestyle{plain}

\noindent\centerline{\Large \bf Harnessing Dark Energy in Five-dimensional }\\ \centerline{\Large \bf Brans-Dicke Universe }\\

\noindent\centerline{$^1$Koijam Manihar Singh}\\
\noindent\centerline{$^2$Kangujam Priyokumar Singh}\\
\textbf{}
\noindent\centerline{$^1$School of Engineering and Technology, Mizoram University}\\
\noindent\centerline{Aizawl, PIN-796004 India}\\
\noindent\centerline{drmanihar@rediffmail.com}\\
\noindent\centerline{Tel.: +918974000569}\\
\noindent\centerline{$^2$Department of Mathematical Sciences, Bodoland University,}\\
\noindent\centerline{Kokrajhar, Assam-783370, India.}\\
\noindent\centerline{pk\_mathematics@yahoo.co.in}\
\noindent\centerline{Tel.: +919856134748}\\
\\
\footnote{Preprint of an article submitted for published in [Canadian Journal of Physics][Journal URL: www.nrcresearchpress.com/journal/cjp]}
\noindent{\bf Abstract.} In trying to explain the present accelerated expansion of the universe in the light of a five-dimensional Brans-Dicke theory, it is found that the fifth dimension itself here acts as a source of dark energy. It may be taken as a curvature-induced form of dark energy, in one case of which it behaves similar to that form of dark energy arising out of the cosmological constant which is the most commonly accepted form of dark energy. It is also found that this new type of dark energy is free from big rip singularity, and may be taken as a viable form of dark energy which can explain some of physical mysteries of the universe.
\vspace{0.2in}

\section{Introduction}
Using extra dimensions beyond the existing four-dimensional space-time, some researchers investigated for new gravitational theories unifying gravity with other forces of nature. Mention may be made of Nordstrom (1914) who investigated for a unified theory based on extra dimensions, and Kaluza (1921) and Klein (1926) theory in which a five-dimensional relativity theory was established in which an extra fifth dimension gives rise to electrodynamics. Some other researches are also found to be performed on similar foundations either via different mechanism of compactification of extra dimension or generalizing it to non-compact scenerios [Overduin and Wesson (1997)] such as brane-world theories (Pavsic (2006), gr-qc/0610061),the space-time-matter or induced-matter (IM) theories [Wesson (1999), Wesson (2006)], and higher dimensional cosmology [Leon (2011)], Compbell-Magaard theorem has great significance on these theories as the theorem asserts that every analytical N-dimensional Riemannian space can be locally embedded in an (N+1)-dimensional vacuum one [Compbell (1926), Magaard (1963), Romero et al.(1995), Lidsey et al.(1997), Seahra and Wesson (2003)]. It can be inferred from this theorem that matter sources of 4D space- times can be taken as a manifestation of extra dimensions. Thus 5D field equations without matter sources can be reverted back to 4D field equations with matter sources. In this way the extra-dimension itself can manifest as some kind of matter source. In compactification of extra curved dimensions it can be expected to have a higher dimensional energy- momentum tensor or a scalar field. It has been obtained by some workers, namely Singh et al. (hep-th/0206193), Vishwakarama and Singh (2003), that a geometrically originated dark energy caused by an addition of a brane curvature scalar term in the action can explain the present accelerated expansion of the universe. Aguilar et al.(2008) showed that 5D vacuum BD equations, when reduced to four dimensions give a modified version  of the four-dimensional BD theory with an induced potential, by making investigation on a 5D vacuum Brans-Dicke theory based on the idea of induced matter theory. It is sometimes desirable to have a higher dimensional energy-momentum tensor or a scalar field , for example, in compactification of the extra curved dimensions [Appelquist(1987)], even though one of the aims of higher dimensional theories is to obtain 4D matter from pure geometry. In the light of induced-matter theory, Bahrehbakhsh et al.(2011) studied the cosmological implications of a five-dimensional Brans -Dicke theory, and tried to explain the acceleration of the universe.\\
\\
Dark energy, which is assumed to be the cause of the present accelerated expansion of the universe, was studied in different forms by different workers. The important forms of dark energy investigated are quintessence, K-essence, f-essence, cosmological constant, tachyon, phantom, chaplygin gas and its generalisations, branes, cosmological nuclear energy, quinton, chameleon, holographic dark energy. And some of the works that(worth mentioning) may be mentioned in these fields are those of Sen and Seshadri (2003), Chimanto and Feinsteim (2004), Myrzakulov (2011), Sahni and Starobinsky (2000), Ellis (2003), Padmanabhan (2002a,b),  Sen (2003,2005), Bangla et al.(1996), Sen et al.(2001), Sen and Scherrer(2005), Caldwell(2002), Nojiri et al.(2005), Jassal(2003), Sahni et al. (2008), Gupta and Pradhan (2010), Adhav et al. (2011a,b), Reddy et al. (2012, 2013), Pradhan (2013), Sahoo and Mishra (2014), Rao et al.(2012), Rao and Neelima (2013), Katore et al.(2012), Saha and Yadav (2012), Samanta (2013), Samanta et al.(2013,2014), Misra and Sahoo(2014), Katore et al.(2011), Bali and Singh (2012), Mahanta et al.(2014), studied the different candidates of dark energy. Manihar and Priyokumar (2014) obtained a model universe consisting partly of quintessence form of the dark energy and partly of cosmological constant form of dark energy and studied the present state of universe. In one of their papers (Manihar et al,2015a) they obtained the possibility of cosmic time going back in some universe dominated by dark energy. Manihar and Priyokumar (2015), Manihar et al.(2015b)obtained also some models of the universe containing dark energy with interesting results and properties. Manihar and Mahanta (2016)obtained as a result of their investigation(study) that there is high possibility of the energy produced from the neutrinos might constitute to the dark energy prevalent in this universe.\\
\\
Here considering a five-dimensional Brans-Dicke theory in the Jordan frame, we study the contribution of the fifth dimension to our traditional four-dimensional universe. Though the fifth dimension seems to shrink with time, it is found that it acts as a source of dark energy which plays a great role in producing the accelerated expansion of the universe. This may be treated as a form of curvature-induced energy which is already accepted and worked upon by some investigators, and we get some interesting results from such types of interactions.\\
\section{Field Equations And Their Solutions}
\label{sect:Obs}
We define the action of five-dimensional BD theory in the Jordan frame as
\begin{equation}
S[g_{AB},\phi]=\int\sqrt{g_{(5)}}(\phi_{(5)}R-\frac{\omega}{\phi}g^{AB}\phi_{,A}\phi_{,B}+16\pi L_{m})d^{5}x
\end{equation}

Where c=1, A,B=0,1,2,3,4 ,R is the five dimensional Ricci scalar, $g_{(5)}$  being the determinant of the metric coefficients $g_{AB}$. Here $\phi$ represents a positive scalar field that describes the gravitational coupling in five dimensions, $L_{m}$ being the matter Lagrangian and a dimensionless coupling constant.
Now from (1) we get the field equations
\begin{equation}
 G_{AB_{(5)}}=\frac{8\pi}{\phi}T_{AB_{(5)}}+\frac{\omega}{\phi^{2}}(\phi_{,A}\phi_{,B}-\frac{1}{2}g_{AB}\phi^{,c}\phi_{,c})+\frac{1}{\phi}(\phi_{;AB}-g_{AB} \square\phi_{(5)})
\end{equation}
 and
\begin{equation}
\square\phi_{(5)}=\frac{8\pi}{4+3\omega}T_{(5)}~~~~~~~~~~~~~~
\end{equation}
 where $G_{AB_{(5)}}$ is five dimensional Eienstein tensor, $T_{AB_{(5)}}$ is the five-dimensional energy-momentum tensor. Here, $T\equiv T_{c}^{c}$ and $\ \Box\phi\equiv \phi_{;A}^{A}$. In this case $T_{AB_{(5)}}$ represents the baryonic and dark matter sources of a four-dimensional hypersurface. Thus here it can been seen that the energy-momentum tensor $T_{AB_{(5)}}$ = $diag(\rho_{M},-p_{M}, -p_{M}, -p_{M},0)$ where $\rho_{M}$ and $p_{M}$ are respectively the energy density and the pressure of matter.Here we consider the five-dimensional metric
\begin{multline}
ds^{2}=g_{AB}(x^{c})dx^{A}dx^{B}=g_{\mu\nu_{(5)}}(x^{c})dx^{\mu}dx^{\nu}+g_{44}(x^{c})dy^{2}
\\ \equiv g_{\mu\nu_{(5)}}(x^{c})dx^{\mu}dx^{\nu}+\epsilon b^{2}(x^{^{c}})dy^{2}~~~~~~~~~~~~~~~~~~~
 \end{multline}
 in the co-ordinate system $x^{A}=(x^{i},y)$, where y represents the fifth coordinate and $\epsilon^{2}=1$. Making the assumption that 5D space-time foliated by a family of hypersurfaces, say $\sum$, which are defined by fixed values of y, we can get the metric intrinsic to every generic hypersurface, for example, $\sum_{0}(y=y^{0})$ by restricting the line element (4) to displacements confined to it.In this way the induced metric on the hypersurface $\sum_{0}$ can be taken as
 \begin{equation}
 ds^{2}=g_{\mu\nu_{(5)}}(x^{\alpha},y_{0})dx^{\mu}dx^{\nu}\equiv g_{\mu\nu}dx^{\mu}dx^{\nu}
 \end{equation}
 so that in this way the usual 4D space-time metric $g_{ij}$ can be recovered.
 Then equation (2) on the hypersurface $\sum_{0}$ takes the form
 \begin{equation}
 G_{\alpha\beta}=\frac{8\pi}{\phi}[T_{\alpha\beta}^{(H)}+T_{\alpha\beta}^{(D)}]
 \end{equation}
 $T_{\alpha\beta}^{(D)}$ being the dark energy component of the energy-momentum tensor which is defined as
\begin{multline}
T_{\alpha\beta}^{(D)}\equiv T_{\alpha\beta}^{(IM)}+T_{\alpha\beta}^{(\phi)}+\frac{\omega}{8\pi\phi}(\phi_{,\alpha}\phi_{,\beta}-\frac{1}{2}g_{\alpha\beta}\phi^{,\sigma}\phi_{,\sigma}) \\ +\frac{1}{8\pi}(\phi_{;\alpha\beta}-g_{\alpha\beta}\Box\phi)~~~~~~~~~~~~~~~~~~~~~~~~~~~~~~~~~~~~~~~~~~~~
\end{multline}
where we take, analogous to the IM theory,
\begin{multline}
T_{\alpha\beta}^{(IM)}\equiv\frac{\phi}{8\pi}\{\frac{b_{;\alpha\beta}}{b}-\frac{\Box b}{b}g_{\alpha\beta}-\frac{\epsilon}{2b^{2}}[\frac{b^{\prime}}{b}g_{\alpha\beta}^{\prime}-g_{\alpha\beta}^{\prime\prime}+g^{\mu\nu}g_{\alpha\mu}^{\prime}g_{\beta\nu}^{\prime}
\\-\frac{1}{2}g^{\mu\nu}g_{\mu\nu}^{\prime}g_{\alpha\beta}^{\prime} -g_{\alpha\beta}(\frac{b^{\prime}}{b}g^{\mu\nu}g_{\mu\nu}^{\prime}-g^{\mu\nu}g_{\mu\nu}^{\prime\prime}-\frac{1}{4}g^{\mu\nu}g^{ij}g_{\mu\nu}^{\prime}g_{ij}^{\prime}-\frac{3}{4}g'^{ij}g'_{ij})]\}
\end{multline}
And
\begin{equation}
T_{\alpha\beta}^{(\phi)}\equiv-\frac{\epsilon}{8\pi b^{2}}[g_{\alpha\beta}\{\phi^{\prime\prime}+(\frac{1}{2}g^{\mu\nu}g_{\mu\nu}^{\prime}-\frac{b'}{b}+\frac{\omega\phi'}{2\phi})\phi'+\epsilon bb_{,\mu}\phi^{,\mu}\}-\frac{1}{2}g'_{\alpha\beta}\phi']
\end{equation}
      where the prime denotes the derivative with respect to the fifth coordinate.
      Here in our problem we consider a generalized five-dimensional universe with the metric
\begin{equation}
ds^{2}=-dt^{2}+a^{2}(t)[\frac{dr^{2}}{1-kr^{2}}+r^{2}(d\theta^{2}+sin^{2}\theta d\phi^{2})]+b^{2}(t)dy^{2}
\end{equation}
Considering the physical feasibility it is assumed that the hypersurface - orthogonal space-like is a killing vector field in the underlying five-dimensional space-time.In the compactified extra dimension scenarios, all fields can be Fourier -expanded around the fixed value. $y_{0}$, thereby getting the observable terms independent of y which means that the physics would be independent of the compactified fifth dimension.
    Here we obtain the Brans-Dicke equations to the metric (10) as
\begin{multline}
(\frac{\dot{a}}{a})^{2}=\frac{8\pi}{3\phi}\rho_{M}+\frac{\omega}{6}(\frac{\dot{\phi}}{\phi})^{2}-\frac{\dot{a}\dot{b}}{ab}+\frac{\ddot{\phi}}{3\phi}-\frac{k}{a^{2}} \\ \text{with}~~~~  \frac{8\pi}{3\phi}\rho_{M}+\frac{\omega}{6}(\frac{\dot{\phi}}{\phi})^{2}-\frac{\dot{a}\dot{b}}{ab}+\frac{\ddot{\phi}}{3\phi}-\frac{k}{a^{2}}=\frac{8\pi}{3\phi}(\rho_{M}+\rho_{D})-\frac{k}{a^{2}}~~~~~~~~~
\end{multline}
\begin{multline}
(\frac{\ddot{a}}{a})=-\frac{4\pi}{\phi}p_{M}-\frac{1}{2}(\frac{\dot{a}}{a})^{2}-\frac{\omega}{4}(\frac{\dot{\phi}}{\phi})^{2}+\frac{1}{2}(\frac{\dot{a}\dot{\phi}}{a\phi})-\frac{\dot{a}\dot{b}}{ab}-\frac{\ddot{b}}{2b}-\frac{k}{2a^{2}}
\\ \text{with} -\frac{4\pi}{\phi}p_{M}-\frac{1}{2}(\frac{\dot{a}}{a})^{2}-\frac{\omega}{4}(\frac{\dot{\phi}}{\phi})^{2}+\frac{1}{2}(\frac{\dot{a}\dot{\phi}}{a\phi})-\frac{\dot{a}\dot{b}}{ab}-\frac{\ddot{b}}{2b}-\frac{k}{2a^{2}} \\ =-\frac{4\pi}{3\phi}[(\rho_{M}+\rho_{D})+3(p_{M}+p_{D})]~~~~~~~~~~~~~~~~~~~~~~~
\end{multline}
\begin{equation}
\frac{\ddot{a}}{a}=-(\frac{\dot{a}}{a\dot{}})^{2}-\frac{\omega}{6}(\frac{\dot{\phi}}{\phi})^{2}+\frac{1}{3}(\frac{\dot{b}}{b})(\frac{\dot{\phi}}{\phi})-\frac{k}{a^{2}}
\end{equation}
Also (3) gives
\begin{equation}
\frac{\ddot{\phi}}{\phi}+3(\frac{\dot{a}}{a})(\frac{\dot{\phi}}{\phi})=\frac{8\pi(\rho_{M}-3p_{M_{}})}{(4+3\omega)\phi}-\frac{\dot{b}}{b}\frac{\dot{\phi}}{\phi}
\end{equation}
  And from (11) $\&$ (12), we see that the energy density $\rho_{D}$ and pressure $p_{D}$ of dark energy as
\begin{equation}
\rho_{D}=\frac{\phi}{8\pi}[\frac{\omega}{2}(\frac{\dot{\phi}}{\phi})^{2}-3(\frac{\dot{a}}{a})(\frac{\dot{b}}{b})+\frac{\ddot{\phi}}{\phi}]
\end{equation}
\begin{equation}
p_{D}=\frac{\phi}{8\pi}[\frac{\omega}{2}(\frac{\dot{\phi}}{\phi})^{2}-(\frac{\dot{a}}{a})(\frac{\dot{\phi}}{\phi})+2(\frac{\dot{a}}{a})(\frac{\dot{b}}{b})+\frac{\ddot{b}}{b}]
\end{equation}
Thus here the equation of state of dark energy is obtained as
\begin{equation}
\omega_{D}=\frac{p_{D}}{\rho_{D}}=\frac{\frac{\omega}{2}(\frac{\dot{\phi}}{\phi})^{2}-(\frac{\dot{a}}{a})(\frac{\dot{\phi}}{\phi})+2(\frac{\dot{a}}{a})(\frac{\dot{b}}{b})+\frac{\ddot{b}}{b}}{\frac{\omega}{2}(\frac{\dot{\phi}}{\phi})^{2}-3(\frac{\dot{a}}{a})(\frac{\dot{b}}{b})+\frac{\ddot{\phi}}{\phi}}
\end{equation}
And the energy conservation equation for the dark energy gives
\begin{equation}
\dot{\rho}_{D}+3(\frac{\dot{a}}{a})(\rho_{D}+p_{D})=(\frac{\dot{\phi}}{\phi})\rho_{D}-Q
\end{equation}
Q being the interacting terms among matter and dark energy
Q being the interaction term between matter(including dark matter) and dark energy.
Now equation (13) can be separated into two parts as
\begin{equation}
\frac{\ddot{a}}{a}+(\frac{\dot{a}}{a})^{2}+\frac{k}{a^{2}}=z
\end{equation}
And
\begin{equation}
-\frac{\omega}{6}(\frac{\dot{\phi}}{\phi})^{2}+\frac{1}{3}(\frac{\dot{b}}{b})(\frac{\dot{\phi}}{\phi})=-z
\end{equation}
Where z is a separation constant.
In the limiting case when z=0 we get
\begin{equation}
a=(ct-kt^{^{2}})^{\frac{1}{2}}
\end{equation}
And
\begin{equation}
\phi=b_{0}b^{\frac{2}{\omega}}
\end{equation}
Where c and $b_{0}$ are arbitrary constants.
Thus from (11) we obtain
\begin{equation}
\frac{8\pi}{3\phi}\rho_{M}=(\frac{\dot{a}}{a})^{2}-\frac{\omega}{6}(\frac{2\dot{b}}{\omega b})^{2}+(\frac{\dot{a}}{a})(\frac{\dot{b}}{b})-\frac{2}{3\omega}(\frac{2}{\omega}-1)(\frac{\dot{b}}{b})^{2}-\frac{2}{3\omega}(\frac{\ddot{b}}{b})+\frac{k}{a^{2}}
\end{equation}
And (12)gives
\begin{equation}
\frac{4\pi}{\phi}p_{M}=\frac{1}{2}(\frac{\dot{a}}{a})(\frac{2\dot{b}}{\omega b})-\frac{\ddot{a}}{a}-\frac{1}{2}(\frac{\dot{a}}{a})^{2}-\frac{\omega}{4}(\frac{2\dot{b}}{\omega b})^{2}-(\frac{\dot{a}}{a})(\frac{\dot{b}}{b})-\frac{\ddot{b}}{2b}-\frac{k}{2a^{2}}
\end{equation}
Now making use of (23) and (24) in (14) we get
\begin{equation}
b=[(\frac{2+\omega}{\omega})\{4k^{\frac{1}{2}}c_{1}c^{-2}(t-\frac{c}{2k})(\frac{ct}{k}-t^{2})^{\frac{-1}{2}}+c_{2}\}]^{\frac{\omega}{\omega+2}}
\end{equation}
Again
\begin{equation}
\omega_{D}=\frac{(\frac{2}{\omega}+1)(\frac{\dot{b}}{b})^{2}+2(1-\frac{1}{\omega})(\frac{\dot{a}}{a})(\frac{\dot{b}}{b})+\frac{d}{dt}(\frac{\dot{b}}{b})}{(\frac{2}{\omega}+\frac{4}{\omega^{2}})(\frac{\dot{b}}{b})^{2}-3(\frac{\dot{a}}{a})(\frac{\dot{b}}{b})+\frac{2}{\omega}\frac{d}{dt}(\frac{\dot{b}}{b})}
\end{equation}
Thus we see that
\begin{equation}
\omega_{D}=\frac{\omega}{2} ~~for~~  \omega=-2,~~ \frac{2}{3}~~~~~~~
\end{equation}
Therefore  ~~~~$\omega_{D}=-1$ when  $\omega=-2$ \\
     and  ~~~~~~$\omega_{D}=\frac{1}{3}$ ~` when  ~~$\omega=\frac{2}{3}$
\\ Again in this problem
\begin{equation}
\phi=b_{0}[(\frac{2+\omega}{\omega})\{4k^{\frac{1}{2}}c_{1}c^{-2}(t-\frac{c}{2k})(\frac{ct}{k}-t^{2})^{\frac{-1}{2}}+c_{2}\}]^{\frac{2}{\omega+2}}
\end{equation}
Also we see that,
\\ The deceleration parameter due to the fifth dimension,
\begin{multline}
q_{b}=-1-c_{1}^{-1}k^{\frac{3}{2}}(\frac{\omega}{\omega+2})^{-1}t^{\frac{3}{2}}(\frac{c}{k}-t)^{\frac{3}{2}}[3\{4k^{\frac{1}{2}}c_{1}c^{-2}(t-\frac{c}{2k})\times
\\(\frac{ct}{k}-t^{2})^{-\frac{1}{2}}+c_{2}\}(t-\frac{c}{2k})(\frac{ct}{k}-t^{2})^{-1}-c_{1}k^{-\frac{3}{2}}]~~~~~~~~~~~~~~~~
\end{multline}
Here the expansion factor due to the fifth dimension,
\begin{equation}
\theta=3c_{1}(c-kt)^{-\frac{3}{2}}(\frac{\omega}{\omega+2})\{4c_{1}c^{-2}t(kt-\frac{c}{2})(c-kt)^{-\frac{1}{2}}+c_{2}t^{\frac{3}{2}}\}^{-1}(c-kt)^{-\frac{3}{2}}
\end{equation}
Also here
\begin{multline}
p_{D}=-\frac{5}{6}c_{1}k^{-3}(\frac{\omega}{\omega+2})(\frac{c}{k}-2t)(\frac{ct}{k}-t^{2})^{-\frac{5}{2}}\times [4k^{\frac{1}{2}}c_{1}c^{-2}(t-\frac{c}{2k}) \\ (\frac{ct}{k}-t^{2})^{-\frac{1}{2}}+c_2]^{-1}-\frac{2(9-\omega^{2})}{9(\omega+2)^{2}}c_{1^{2}}k^{-3}(\frac{ct}{k}-t^{2})^{-3}\times \\ [4k^{\frac{1}{2}}c_{1}c^{-2}(t-\frac{c}{2k})(\frac{ct}{k}-t^{2})^{-\frac{1}{2}}+c_2]^{-2}~~~~~~~~~~~~~~~~~~~~~~~
\end{multline}
\begin{multline}
\rho_{D}=\frac{b_{0}}{8\pi}(\frac{\omega+2}{\omega})^{\frac{2}{\omega+2}}\{4k^{\frac{1}{2}}c_{1}c^{-2}(t-\frac{c}{2k})(\frac{ct}{k}-t^{2})^{-\frac{1}{2}}+c_{2}\}^{-\frac{2(\omega+1)}{\omega+2}}\times \\ [12c_{1}^{2}c^{-2}k^{-1}(t-\frac{c}{2k})^{2}(\frac{ct}{k}-t^{2})^{-3}+3c_{1}c_{2}k^{-\frac{3}{2}}(t-\frac{c}{2k})(\frac{ct}{k}-t^{2})^{-\frac{5}{2}}]
\end{multline}
And
\begin{multline}
Q=\frac{b_{0}^{2}}{8\pi}c_{1}k^{-\frac{3}{2}}(\frac{2}{\omega+2})(\frac{\omega+2}{\omega})^{\frac{2}{\omega+2}}(\frac{ct}{k}-t^{2})^{-\frac{3}{2}}\{4k^{\frac{1}{2}}c_{1}c^{-2}(t-\frac{c}{2k}) \\ (\frac{ct}{k}-t^{2})^{-\frac{1}{2}}+c_{2}\}^{-\frac{(3\omega+4)}{\omega+2}}\{12c_{1}^{2}c^{-2}k^{-1}(t-\frac{c}{2k})^{2}(\frac{ct}{k}-t^{2})^{-3}
\\ +3c_{1}c_{2}k^{-\frac{3}{2}}(t-\frac{c}{2k})(\frac{ct}{k}-t^{2})^{-\frac{5}{2}}\}+\frac{b_{0}}{4\pi}c_{1}k^{-\frac{3}{2}}(\frac{\omega+2}{\omega})^{\frac{2}{\omega+2}}
\\(\frac{\omega+1}{\omega+2})(\frac{ct}{k}-t^{2})^{-\frac{3}{2}}\{4k^{\frac{1}{2}}c_{1}c^{-2}(t-\frac{c}{2k})(\frac{ct}{k}-t^{2})^{-\frac{1}{2}}+c_{2}\}^{-\frac{(3\omega+4)}{\omega+2}}
\\ \{12c_{1}^{2}c^{-2}k^{-1}(t-\frac{c}{2k})^{2}(\frac{ct}{k}-t^{2})^{-3}+3c_{1}c_{2}k^{-\frac{3}{2}}(t-\frac{c}{2k})(\frac{ct}{k}-t^{2})^{-\frac{5}{2}}\}
\\-\frac{b_{0}}{8\pi}(\frac{\omega+2}{\omega})^{\frac{2}{\omega+2}}\{4k^{\frac{1}{2}}c_{1}c^{-2}(t-\frac{c}{2k})(\frac{ct}{k}-t^{2})^{-\frac{1}{2}}+c_{2}\}^{-\frac{2(\omega+1)}{\omega+2}}
\\ \{24c_{1}^{2}c^{-2}k^{-1}(t-\frac{c}{2k})(\frac{ct}{k}-t^{2})^{-3}+72c_{1}^{2}c^{-2}k^{-1}(t-\frac{c}{2k})^{3}(\frac{ct}{k}-t^{2})^{-4}+
\\3c_{1}c_{2}k^{-\frac{3}{2}}(\frac{ct}{k}-t^{2})^{-\frac{5}{2}}+15c_{1}c_{2}k^{-\frac{3}{2}}(t-\frac{c}{2k})^{2}(\frac{ct}{k}-t^{2})^{-\frac{7}{2}}\}+\frac{3kb_{0}}{8\pi}
\\(\frac{\omega+2}{\omega})^{\frac{2}{\omega+2}}(t-\frac{c}{2k})(ct-kt^{2})^{-1}\{4k^{\frac{1}{2}}c_{1}c^{-2}(t-\frac{c}{2k})(\frac{ct}{k}-t^{2})^{-\frac{1}{2}}+
\\c_{2}\}^{-\frac{2(\omega+1)}{\omega+2}}\{12c_{1}^{2}c^{-2}k^{-1}(t-\frac{c}{2k})^{2}(\frac{ct}{k}-t^{2})^{-3}+3c_{1}c_{2}k^{-\frac{3}{2}}(t-\frac{c}{2k})
\\(\frac{ct}{k}-t^{2})^{-\frac{5}{2}}\}+5c_{1}k^{-3}(\frac{\omega}{\omega+2})(t-\frac{c}{2k})^{2}(\frac{ct}{k}-t^{2})^{-\frac{7}{2}}\{4k^{\frac{1}{2}}c_{1}c^{-2}
\\(t-\frac{c}{2k})(\frac{ct}{k}-t^{2})^{-\frac{1}{2}}+c_2\}^{-1}-\frac{2}{3}c_{1}^{2}k^{-3}\frac{(9-\omega^{2})}{(\omega+2)^{2}}(t-\frac{c}{2k})(\frac{ct}{k}-t^{2})^{-4}
 \\ \{4k^{\frac{1}{2}}c_{1}c^{-2}(t-\frac{c}{2k})(\frac{ct}{k}-t^{2})^{-\frac{1}{2}}+c_{2}\}^{-2}~~~~~~~~~~~~~~~~~~~~~~
\end{multline}

\section{Study of the Solution and Conclusion}
\label{sect:Interpretation}

From the results obtained here it is seen that the fifth dimension in the Brans-Dicke field plays a great role in the evolution of the universe and in causing the present accelerated expansion of the universe.If the Brans-Dicke coupling constant happens to be $\frac{2}{3}$ then the scale factor of the fifth dimension along with the scalar field play a great role in the producing a radiation-dominated universe as it was in the  previous era.At the present epoch, the equation of state $\omega_{D}=\frac{p_{D}}{\rho_{D}}$ comes out to be -1 corresponding to the coupling constant $\omega=-2$, which indicates that the scale factor of the five dimension contributes to the so-called dark energy, the form being of the cosmological constant type.But as $\omega\rightarrow-2$ we see that the scale factor of the five dimension $b\rightarrow0$, and it symbolizes the present scenario of the universe.Thus though it contributes to the dark energy, the fifth dimension shrinks and it is almost invisible at the present epoch.The expansion factor $\theta$  due to the fifth dimension increases with time, and the deceleration parameter comes out to be a negative quantity which bears testimony to the present accelerated expansion of the universe.Again it seems that as $\omega$ increases from $\frac{2}{3}$ to -2, the value of $\omega_{D}$ gradually decreases from $\frac{1}{3}$ to -1, which indicates the solution of a realistic universe passing through radiation and matter dominated phases, then coming to that of a dark energy dominated phase as is at present.In this problem it is also seen that the fifth dimension activates the scalar field which, in turn, helps in enhancing the quantum of producing dark energy.And also as a particular case, if $b_{0}=0$ we find that the scalar field tends to zero and in such a situation it can be conveniently taken that for such a universe the origin of dark energy is due to the scale factor of the fifth dimension.As at the present epoch when the dark energy dominates the universe the fifth dimension shrinks or compactifies to an infitesimally small quantity, thus the length of the present time from the beginnings of the universe is given by the value of t in the equation\\ \\
\noindent\centerline{
 \text~~~~~~~~~~~~~~~~~~~~$4(t-\frac{c}{2k})=k^{-\frac{1}{2}}c^{2}c_{1}^{-1}c_{2}(\frac{c}{k}t-t^{2})^{\frac{1}{2}}$}
\\ \\ \text~~~~~~~As seen here the field due to the fifth dimension is seen to behave similar to the Brans-Dicke scalar field.Also the pressure due to the fifth dimension here is found to be negative which strengthens our assumption that the fifth dimension , though invisible now, contributes as a source of dark energy in the present universe.Further it is seen that there will not be big rip singularities due to this form of dark energy.
\\

\end{document}